\documentclass {article}
\usepackage [latin1]{inputenc}
\usepackage{amsmath, amsthm, amssymb}[]
\usepackage{dsfont}
\usepackage{bm}
\newcommand{\funi}{\footnotesize\textbf{i\hspace{-0.22
cm}\normalsize$\bigcirc$
}}

\newcommand{\funcii}{\footnotesize\textbf{i\hspace{-0.22
cm}\normalsize$\bigcirc_{i}$}
}
\newcommand{\funcip}{\footnotesize\textbf{i\hspace{-0.22
cm}\normalsize$\bigcirc_{\mathcal{P}_{\bm{t}}}$}
}
\newcommand{\funcipu}{\footnotesize\textbf{i\hspace{-0.22
cm}\normalsize$\bigcirc_{\mathcal{P}_{\bm{t+1}}}$}
}
\newcommand{\funcipd}{\footnotesize\textbf{i\hspace{-0.22
cm}\normalsize$\bigcirc_{\mathcal{P}_{\bm{t+2}}}$}
}

\begin{document}
\title{Multi-optional Many-sorted Past Present Future structures  and its description}
\author{Sergio Miguel Tom\'e\\ \\
\small{Grupo de Investigaci\'on en Miner\'ia de Datos (MiDa)},\\
\small{Universidad de Salamanca, Salamanca, Spain}\\
\small{sergiom@usal.es}}

\maketitle
\begin{abstract}
The cognitive theory of true conditions (CTTC) is a proposal to describe the model-theoretic semantics of symbolic cognitive architectures and design the implementation of cognitive abilities. The CTTC is formulated mathematically using the multi-optional many-sorted past present future(MMPPF) structures.  This article defines mathematically the MMPPF structures and the formal languages proposed to describe them by the CTTC.
\end{abstract}

\section{Introduction}
Semantics is one of the most challenging aspects of cognitive architectures. The Cognitive Theory of True Conditions (CTTC) is a proposal to describe the model-theoretic semantics of symbolic cognitive architectures and to develop decision-making processes based on model-theoretic semantics \cite{Sergio2006,SergioTesis}.
The main idea behind the CTTC is that the perceptual space is a set of formal languages that denote elements of a model embedded in a quotient space of the physical space. At this moment, the mathematical formulation of the CTTC is using the \emph{multi-optional many-sorted past present future}(MMPPF). structures. Also, the CTTC proposes a hierarchy of three formal languages to describe them.

This article improves the previous characterization of the MMPPF structures and the hierarchy of the formal languages \cite{Sergio2006,SergioTesis}. The article  is divided in three sections. The first section gives the mathematical definitions of  MMPPF structures. The second section defines a hierarchy of three formal languages to describe an MMPPF structure. The last section addresses the semantics of the three formal languages of the hierarchy.

\section{ Multi-optional Many-sorted Past Present Future structures }

A MMPPF structure is a nested structure of possible worlds.  In other words, each world of the structure also contains another possible worlds structure. Thus, they are more complex than the classical structures of possible worlds used in temporal logics. This section  includes the definitions of the MMPPF structure, temporal perspective structure, and state structure and its axioms. An MMPPF structure is constructed using temporal perspectives structures, and a temporal perspective structure is constructed using state structures. Thus, we define firstly a state structure, after a temporal perspective structure and finally an MMPPF structure. After the definitions, the axioms of the MMPPF structure are provided.
\paragraph{}

A \emph{state structure} is a many-sorted structure, and it is denoted by $\bm{e}$.  Its definition  is the following:

\[
\bm{e}= \langle \langle\bm{U}_{i}\rangle \langle \bm{f}_{l} \rangle \langle \bm{R}_{k}\rangle   \rangle
\]
where the domains are

\[
\langle \bm{U}_{i}\rangle = \langle \mathcal{\bm{O}}, \mathcal{P}(\bm{H}),\mathcal{P}(\bm{H}\times \bm{V}_{0}),...,\mathcal{P}(\bm{H}\times \bm{V}_{n}), \bm{A}^{0}_{1},...,\bm{A}^{n}_{1},...,\bm{A}^{0}_{z},...,\bm{A}^{n}_{z}, \bm{SRA} \rangle
\]
\begin{itemize}
  \item \[
\mathcal{\bm{O}} = \{\bm{o}_{1},...,\bm{o}_{z} \}
\]
  \item \[
\bm{H} = \{ \bm{h}_{1},..., \bm{h}_{z'} \} \quad  z \leq z'
\]
  \item \[
\bm{V}_{p} = \{ \emptyset \} \cup \{ (\bm{w}_{1},...,\bm{w}_{dim(p)}): \bm{w}_{1} \in \bm{W}_{p,1},...\bm{,w}_{dim(p)} \in \bm{W}_{p,dim(p)} \}
\]

where
\[
\begin{array}{cccc}
  dim :& P & \longrightarrow & \mathds{N}
\end{array}
\]

  \item \[
\bm{A}^{p}_{\bm{o}_{i}} = \{ \bm{a}^{p}_{\bm{o}_{i},1},...\bm{a}^{p}_{\bm{o}_{i},k}\} \quad  \bm{o}_{i} \in  \mathcal{O} \textrm{ and } \bm{a}^{p}_{\bm{o}_{i},k}=(\bm{a}^{p}_{\bm{o}_{i},k}.in, \bm{a}^{p}_{\bm{o}_{i},k}.ext )
\]
where
\[
  \bm{a}^{p}_{\bm{o}_{i},k}.in: \bm{DH}_{p,k} \times \bm{DV}_{p,k}\longrightarrow \bm{CV} \quad \bm{DH}_{p,k} \subseteq \bm{H} , \; \bm{DV}_{p,k}, \bm{CV}_{p,k} \subseteq \bm{V}_{p}
\]
\[
\bm{a}^{p}_{\bm{o}_{i},k}.ext: \bm{H}_{p,k} \times \bm{DV}_{p,k}\longrightarrow \bm{CV} \quad \bm{H}_{p,k} \subseteq \bm{H} , \; \bm{DV}_{p,k}, \bm{CV}_{p,k} \subseteq \bm{V}_{p}
\]

  \item \[
\bm{SRA }= \{\langle(\bm{s}_{1},\bm{c}_{1}),....,(\bm{s}_{r},\bm{c}_{r})\rangle,...,\langle(\bm{c}_{1},\bm{c}_{1}),....,(\bm{c}_{r},\bm{c}_{r})\rangle \} \cup \{ \emptyset\}
\]
\end{itemize}

the functions are
\begin{itemize}
  \item \[
\langle \bm{f}_{l}\rangle = \langle \bm{ ES}_{j},\bm{g}^{*0}_{j},...,\bm{g}^{*n}_{j},\bm{g}^{0}_{j},...,\bm{g}^{n}_{j} ,\theta_{0}^{p},...,\theta_{n}^{p}, \odot \rangle
\]
  \item \[
\bm{ES}_{j}: \mathcal{O} \longrightarrow \mathcal{P}(\bm{H})
\]
  \item \[
\bm{g}^{*p}: \mathcal{O}  \longrightarrow \mathcal{P}(\bm{H}\times \bm{V}_{p})
\]

\item
\[
\bm{g}^{p}: \bm{H} \longrightarrow \bm{V}_{p}
\]
\item
\[
\bm{\theta}^{p} : \mathcal{\bm{O}}  \longrightarrow \cup_{i}\mathcal{P}(\bm{A}^{p}_{i})
\]
\item
\[
\odot : \mathcal{\bm{O}} \longrightarrow \bm{SRA}
\]
\end{itemize}

and the relations are

\[
\langle \bm{R}_{k} \rangle= \langle  \bm{S}_{j}^{\bm{o}_{1},0},...,\bm{S}_{j}^{\bm{o}_{1},n},...,\bm{S}_{j}^{\bm{o}_{z},0},...,\bm{S}_{j}^{\bm{o}_{z},n}   \rangle
\]
where
\[
\bm{S}_{j}^{\bm{o}_{i},p} \subseteq \mathcal{\bm{O}}
\]

The definition of an \emph{structure of temporal perspective}, $\mathcal{P}_{\bm{t}}$, is

\[
\mathcal{P}_{\bm{t}} = \langle \bm{M}_{\mathcal{P}_{\bm{t}}},  \bm{T}_{\mathcal{P}_{\bm{t}}}, \bm{E}_{\mathcal{P}_{\bm{t}}} ,\bm{I}_{\mathcal{P}_{\bm{t}}}, \mathcal{\bm{D}}^{p}_{\mathcal{P}_{\bm{t}}} ,  \langle \bm{d}^{p}_{j}\rangle, \bm{L}_{\mathcal{P}_{\bm{t}}}, \bm{SL}_{\mathcal{P}_{\bm{t}}}, \bm{\&}_{\mathcal{P}_{\bm{t}}}, \,\funcip, \bm{Succ}_{\mathcal{P}_{\bm{t}}}, \prec_{\mathcal{P}_{\bm{t}}}  \rangle
\]
where its domains are the following:
\begin{itemize}
  \item \[
\bm{M}_{\mathcal{P}_{\bm{t}}} = \{\bm{m}_{1},....,\bm{m}_{z} \}
\]
Each element of $\bm{M}_{\mathcal{P}_{\bm{t}}}$ is denominated moment of time. Each moment of time is a set defined in the following way:
\[
\bm{m}_{t'} \subset  \bm{T}_{\mathcal{P}_{\bm{t}}}\times\complement \times\mathfrak{S}\times \bm{E}_{\mathcal{P}_{\bm{t}}} \quad \complement = \{ \bm{h}, \bm{\varepsilon} \}  \quad  \mathfrak{S} = \{ \downarrow|, ||, |\downarrow \}
\]

 Being $\bm{t}$ the constant denoted by the temporal perspective  $\mathcal{P}_{\bm{t}}$, the elements of a moment of time, $\bm{m}_{x} \in \bm{M}_{\mathcal{P}_{\bm{t}}}  $ , where $\bm{r} \in \bm{m}_{\bm{t}'}$, $\bm{r} = (\bm{t}',\blacklozenge, \blacksquare, \bm{e}_{y} )$ , $\bm{t}' \in \bm{T}_{\mathcal{P}_{\bm{t}}}$,  $\blacklozenge\in \complement$,   $\blacksquare \in \mathfrak{S}$ and $\bm{e}_{y} \in  \bm{E}_{\mathcal{P}_{\bm{t}}}$, fulfill the following:
\begin{itemize}
  \item If $\bm{t}' < \bm{t} $ then $ \blacksquare = \downarrow|$
  \item If $\bm{t}' = \bm{t}$ then $ \blacksquare = ||$
  \item If $\bm{t}' > \bm{t}$ then $  \blacksquare = |\downarrow $
\end{itemize}

Each element of a moment of time is denominated \emph{reality}.

  \item \[
\bm{T}_{\mathcal{P}_{\bm{t}}}=\{1,...,z\} \textrm{ is the time set }
\]
It must be noted that the number of moment of times is the same to the number of elements that has $\bm{T}$
  \item
\[
\bm{E}_{\mathcal{P}_{\bm{t}}} = \{\bm{e}_{1},...,\bm{e}_{s} \}  \textrm{ is a set of states structures}
\]

\item \[
\bm{I}_{\mathcal{P}_{\bm{t}}} = \bigcup_{j} \bm{I}_{e_{j}} \textrm{ and }  \bm{I}_{e_{j}}= (\theta^{0}_{j}(o_{1})\times\cdots \times \theta^{n}_{j}(o_{1}))\times\cdots \times (\theta^{0}_{j}(o_{z})\times\cdots \times \theta^{n}_{j}(o_{z})) \quad e_{j} \in \bm{E}_{\mathcal{P}_{\bm{t}}}
\]
\end{itemize}

Associated with $\bm{I}_{\mathcal{P}_{\bm{t}}}$ we use an auxiliary function, $\pi^{\bm{o}_{i_{n}}}$, which projects a part of an element  $\vec{\bm{i}}$ that belongs to $\bm{I}_{\mathcal{P}_{\bm{t}}}$ , to do definitions. The function is defined in the following way:
\[
\pi^{\bm{o}_{n}}(\langle\vec{\bm{a}}_{\bm{o}_{1}},...,\vec{\bm{a}}_{\bm{o}_{z}}\rangle) = \vec{\bm{a}}_{\bm{o}_{n}}
\]
where $\vec{\bm{i}} = \langle\vec{\bm{a}}_{\bm{o}_{1}},...,\vec{\bm{a}}_{\bm{o}_{z}}\rangle$.

The functions are the following:
\begin{itemize}
  \item \[
\mathcal{D}^{p}_{\mathcal{P}_{\bm{t}}} = \bigcup_{j} \bm{D}^{p}_{j} \textrm{ and } \bm{D}^{p}_{j} =\{ \bm{S}_{j}^{\bm{o}_{i},p}\bm{o}_{u}: \bm{o}_{u}\in \bm{e}_{j} \}
\]
  \item \[
\bm{L}_{\mathcal{P}_{\bm{t}}} = \{\bm{l}^{0},...,\bm{l}^{n} \}\rangle \textrm{ is the set of environmental laws of } \mathcal{P}_{\bm{t}}.
\]
  \item \[
\bm{l}^{p}: \mathcal{P}(\mathcal{P}(\bm{H}\times \bm{V}_{p})\times \mathcal{O})\times \mathcal{P}(\bm{A}^{p}\times \mathcal{O})\times \mathcal{\bm{D}}^{p} \longrightarrow \mathcal{P}(\mathcal{P}(\bm{H}\times \bm{V}_{p}))
\]
\item\[
\bm{SL}_{\mathcal{P}_{\bm{t}}}= \{sl^{1},...,sl^{r}\}
\]
where
\[
\bm{sl}^{k}:  \bm{SRA} \times \mathcal{\bm{O}} \times  \bm{E}_{\mathcal{P}_{\bm{t}}}  \times \bm{I}_{\mathcal{P}_{\bm{t}}}  \longrightarrow \bm{SRA}
\]
\item\[
 \bm{\&}_{\mathcal{P}_{\bm{t}}}: \bm{E}_{\mathcal{P}_{\bm{t}}}\times \bm{I} \longrightarrow \bm{E}_{\mathcal{P}_{\bm{t}}}\cup\{\perp \}
\]

where
\[
\bm{\&}_{\mathcal{P}_{\bm{t}}} (\bm{e}_{x},\vec{\bm{i}})\left\{
  \begin{array}{ll}
    =\, \perp, & \hbox{$\vec{\bm{i}}\notin \bm{I}_{\bm{e}_{x}}$;} \\
    =\, \bm{e}_{y} \in E_{\mathcal{P}_{\bm{t}}}, & \hbox{$\vec{i}\in I_{\bm{e}_{x}}$.}
  \end{array}
\right.
\]

\item
\[
\funcip : \bm{T}_{t}' \; \longrightarrow \bm{I}  \textrm{ where } \bm{T}_{t}'=\{ \bm{t}': \bm{t}' \in \bm{T}  \textrm{ and } \bm{t}' \leq \bm{t} \}
\]

\item
\[
  Succ_{\mathcal{P}_{\bm{t}}} : (\bm{T}\times \complement \times \mathfrak{\bm{S}} \times \bm{E} ) \times \bm{I}  \longrightarrow  (\bm{T}\times \complement \times \mathfrak{\bm{S}} \times \bm{E} )
\]
\end{itemize}

and the relation $\prec_{\mathcal{P}_{\bm{t}}}$ is defined as follows:
\[
 \prec_{\mathcal{P}_{\bm{t}}} \subset \bm{M}_{\mathcal{P}_{\bm{t}}}\times \bm{M}_{\mathcal{P}_{\bm{t}}}
 \]
Each $ \langle  (t',\blacklozenge, \blacksquare, e_{x} ), (t'',\blacklozenge, \blacksquare, e_{x}) \rangle \in \prec_{\mathcal{P}_{\bm{t}}}$ fulfills that $ t'< t''$.

 \paragraph{}

 An \emph{MMPPF structure} is formally defined as the tuple
 \[
 \langle \mathcal{\bm{U}}, \bm{T}, \bm{E} ,\bm{I}, \mathcal{\bm{D}}^{p}, \rho \scriptscriptstyle _{MMPPF}\textstyle , \langle \bm{d}^{p}_{j}\rangle, \bm{L} , \bm{\&}, \; \funi \rangle
\]
where its domains are
\begin{itemize}
  \item \[
\mathcal{\bm{U}}=\{ \mathcal{P}_{1},...,\mathcal{P}_{m} \} \textrm{  is the set of temporal perspectives }
\]
  \item \[
\bm{T}=\{1,...,m\} \textrm{ is the time set }
\]
  \item \[
\bm{E} = \{\bm{e}_{1},...,\bm{e}_{s} \}  \textrm{ is the set of states}
\]
\item
\[
\bm{I} = \bigcup_{j} \bm{I}_{\bm{e}_{j}} \textrm{ and }  I_{j}= (\bm{\theta}^{0}_{j}(\bm{o}_{1})\times\cdots \times \bm{\theta}^{n}_{j}(\bm{o}_{1}))\times\cdots \times (\bm{\theta}^{0}_{j}(\bm{o}_{z})\times\cdots \times \bm{\theta}^{n}_{j}(\bm{o}_{z})) \quad \bm{e}_{j} \in \bm{E}
\]

\item
\[
\mathcal{D}^{p} = \bigcup_{j} \bm{D}^{p}_{j} \textrm{ and } \bm{D}^{p}_{j} =\{ \bm{S}^{\bm{o}_{i},p}\bm{o}_{u}: \bm{S}^{\bm{o}_{i},p}\bm{o}_{u}\in \bm{e}_{j} \} \textrm{ is the dependecies set }
\]
\end{itemize}

its functions are
\begin{itemize}
  \item \[
\rho_{\scriptscriptstyle _{MMPPF} \textstyle}: \bm{T} \longrightarrow \mathcal{U}
\]
  \item \[
\bm{d}^{p}_{\bm{e}_{j}}:\mathcal{P}(\mathcal{P}(\bm{H}\times \bm{V}_{0})\times \mathcal{P}(\bm{H}\times \bm{V}_{p})\times \mathcal{O})\times \mathcal{P}(\bm{A}^{p}\times \mathcal{O})\longrightarrow \mathcal{D}^{p} \quad \bm{e}_{j} \in \bm{E}
\]
  \item \[
\bm{L} = \{\bm{l}^{0},...,\bm{l}^{n} \}\textrm{ is the set of environmental laws.}
\]
  \item \[
l^{p}: \mathcal{P}(\mathcal{P}(\bm{H}\times \bm{V}_{p})\times \mathcal{O})\times \mathcal{P}(A^{p}\times \mathcal{O})\times \mathcal{D}^{p} \longrightarrow \mathcal{P}(\mathcal{P}(\bm{H}\times \bm{V}_{p}))
\]

 \item\[
 \bm{\&}: \bm{E}\times \bm{I} \longrightarrow \bm{E}\cup\{\perp \}
\]

where
\[
\bm{\&} (\bm{e}_{x},\vec{\bm{i}})\left\{
  \begin{array}{ll}
    =\, \perp, & \hbox{$\vec{\bm{i}}\notin \bm{I}_{\bm{e}_{x}}$;} \\
    =\, \bm{e}_{y} \in \bm{E}, & \hbox{$\vec{\bm{i}}\in \bm{I}_{\bm{e}_{x}}$.}
  \end{array}
\right.
\]

\item \[
 \funi:  \bm{T} \longrightarrow \bm{I}
\]
\end{itemize}

The following axioms define an MMPPF structure:
\begin{itemize}
  \item  \textbf{First Axiom}
  \[
  \forall e_{j} \in E_{\mathcal{P}_{\bm{t}}}\quad  ( g^{p}_{j}(h)\neq \emptyset \wedge p>0 \Leftrightarrow \{h \}\subset ES_{j}(o_{i}) )
  \]
 The first axiom determines that an essence element has assigned a value of any property if and only if is assigned to an object.
  \item  \textbf{Second Axiom}
  \[
  \forall e_{j} \in E_{\mathcal{P}_{\bm{t}}}  \quad (p>0) \wedge  ( g^{p}_{j}(h)\neq \emptyset \Leftrightarrow  g^{0}_{j}(h)\neq \emptyset  )
  \]

  The second axiom determines that a place of the space is assigned to an object if and only if it has assigned any value of any other property.
  \item  \textbf{Third Axiom}
  \[
  \forall e_{j} \in E_{\mathcal{P}_{\bm{t}}} \quad  ( g^{p}_{j}(h)= x \Leftrightarrow  \exists o_{i} (h,x) \in g^{*p}_{j}(o_{i}) )
  \]

  The third axiom determines the relation between  $g^{p}_{j}$ and $g^{*p}_{j}$.
  \item \textbf{Fourth Axiom}
  \[
  \forall e_{j} \in E_{\mathcal{P}_{\bm{t}}} \quad  ( a^{p}_{i,k}\in A^{p}_{i} \Rightarrow DH_{p,k} = \bigcup_{j} ES_{j}(o_{i}) )
  \]

  The fourth axiom ensures that an action acts independently of the assignment  by $ES_{j}$ .

  \item  \textbf{Fifth Axiom}
  \[
  \forall e_{j}, e_{j'} \in E_{\mathcal{P}_{\bm{t}}} \quad ( h \in ES_{j}(o_{i}) \wedge h \in ES_{j'}(o_{u}) \Leftrightarrow u = i )
  \]

  The fifth axiom determines that an essence element is only assigned to an object.
  \item  \textbf{Sixth Axiom}
  \[
  Succ_{\mathcal{P}_{\bm{t}}}(t,e_{j})=e_{j'} \Leftrightarrow d^{p}_{j} ( \langle (g^{*0}_{j}(o_{i}),o_{i})\rangle_{i}, \langle (g^{*p}_{j}(o_{i}),o_{i})\rangle_{i}, \langle (a^{*p}_{i,k},o_{1})\rangle_{i} ) = D^{p}_{j'}
  \]
  The sixth axiom determines that the dependencies set is coherent with the changes from $e_{j}$ to $e_{j'}$.

  \item  \textbf{Seventh Axiom}
    \[
 Succ_{\mathcal{P}_{\bm{t}}}(t,e_{j})=e_{j'} \Leftrightarrow \exists \langle a^{p}_{i,k},o_{i}\rangle_{i} \quad \forall p \;\langle( g^{*p}_{j'}(o_{i}), o_{i} )\rangle_{i} =l^{p}(\langle( g^{*p}_{j}(o_{i}), o_{i} )\rangle_{i} , \langle a^{p}_{i,k},o_{i}\rangle_{i}), D^{p}_{j} )
  \]
    The seventh axiom determines that if state $e_{j'}$ succeed state $e_{j}$, it is because the objects can produce changes that generate $e_{j'}$ from $e_{j}$ .
  \item  \textbf{Eighth Axiom}
  \[
   \forall e \in E_{\mathcal{P}_{\bm{t}}} \quad \theta^{p}(o_{i}) \in \mathcal{P}(A^{p}_{i}) )
  \]
  The eight axiom determines that $\theta^{p}_{j}$ only assigns actions to an object when they modify that object.

  \item \textbf{Ninth Axiom}
  \[
  \begin{array}{c}
    \forall t \in T \quad T_{\mathcal{P}_{\bm{t}}} = T \\
    \forall t \in T \quad E_{\mathcal{P}_{\bm{t}}} = E\\
    \forall t \in T \quad I_{\mathcal{P}_{\bm{t}}} = I\\
    \forall t \in T \quad L_{\mathcal{P}_{\bm{t}}} = L\\
    \forall t \in T \quad \&_{\mathcal{P}_{\bm{t}}} = \& \\
    \forall t \in T \quad \funcip(t) =\; \funi(t)= \;\funcipu(t)= \; \funcipd(t) =\cdots \\
    \forall t \in T \quad \mathcal{D}^{p}_{\mathcal{P}_{\bm{t}}} = \mathcal{D}^{p}\\
    \forall t \in T \quad \langle d^{p}_{j}\rangle_{\mathcal{P}_{\bm{t}}} =\langle d^{p}_{j}\rangle\\
    \forall \mathcal{P}_{\bm{t}}, \mathcal{P}_{t'} \quad |SL_{\mathcal{P}_{\bm{t}}}|= |SL_{\mathcal{P}_{t'}}|\\
    \forall \mathcal{P}_{\bm{t}}, \mathcal{P}_{t'} \quad  sl^{k}_{\mathcal{P}_{\bm{t}}} = sl^{k}_{\mathcal{P}_{t'}}

  \end{array}
  \]
 The ninth axiom determines that domains do not change from one temporal perspective to other.

  \item  \textbf{Tenth Axiom}
\[
  Succ_{\mathcal{P}_{\bm{t}}}( (t',\blacklozenge ,\blacksquare, e_{x} ),i  )  =
         \begin{cases}
        (t'+1,\varepsilon ,\downarrow |, e_{y} ), & \textrm{for } t'+1< t  \textrm{ and }\;  \funcip (t')= i \textrm{ and }\blacklozenge = \varepsilon  \\
        (t'+1,h ,\downarrow |, e_{y} ), & \textrm{for } t'+1< t  \textrm{ and }\;  \funcip (t')\neq i\\
        (t'+1,\varepsilon , ||, e_{y} ), & \textrm{for } t'+1= t  \textrm{ and }\;  \funcip (t')= i\textrm{ and }\blacklozenge = \varepsilon\\
        (t'+1,h ,||, e_{y} ), & \textrm{for } t'+1= t  \textrm{ and }\;  \funcip (t')\neq i_{j}\\
        (t+1,\varepsilon ,|\downarrow, e_{y} ), & \textrm{for }  t'= t \textrm{ and }\;  \funcip (t')= i \textrm{ and } \blacklozenge = \varepsilon \\
        (t+1,h ,|\downarrow, e_{y} ), & \textrm{for }  t'= t \textrm{ and }\; \funcip (t')\neq i \textrm{ and } \blacklozenge = h\\
        (t+1,\varepsilon ,|\downarrow, e_{y} ), & \textrm{for }  t'> t  \textrm{ and } \blacklozenge = \varepsilon \\
        (t+1,h ,|\downarrow, e_{y} ), & \textrm{for }  t'> t  \textrm{ and } \blacklozenge = h
        \end{cases}
 \]
The tenth axiom  determines in what period of time a temporal moment and its reality condition are situated.
  \item  \textbf{Eleventh Axiom}
\begin{equation*}
\begin{split}
\forall e_{j}, e_{j'} \in E_{\mathcal{P}_{\bm{t}}}\; Succ_{\mathcal{P}_{\bm{t}}}( (t,\blacklozenge ,\blacksquare, e_{j} ),i  )  = (t,\blacklozenge' ,\blacksquare', e_{j'} ) \rightarrow \\
 \rightarrow \odot_{j'}(o_{i}) = sl^{k} ( \odot_{j}(o_{i}), o_{i},e_{j} , i   )
\end{split}
\end{equation*}

The eleventh axiom determines that the change of reward and aversion sensation is coherent with the actions that the object carries out.

\end{itemize}

\section{ Formal languages to the MMPPF structures }

The section defines three formal languages to describe an MMPPF structure: the perceptive language, the extended perceptive language and the categorical language. They are denoted $PL_{PPFMM}$, $PL^{*}_{PPFMM}$ and $CL_{PPFMM}$ respectively.

\subsection{The perceptive language of MMPPF }

The elements of the alphabet of the $PL_{MMPPF}$ language are the following symbols:

\begin{itemize}
  \item A constant symbol $o_{i}$ for each element $\bm{o}_{i}$ of $\mathcal{\bm{O}}$
  \item A constant symbol $h_{i}$ for each element of $\bm{h}_{i}$ of $\bm{H}$
  \item A constant symbol $w^{p,q}_{i}$ for each element of $\bm{w}^{p,q}_{i}$ of each $\bm{W}_{p,q}$.
  \item Two constant symbols, $\varepsilon$ and $h$, for the elements of $\complement= \{ \bm{\varepsilon}, \bm{h} \}$
  \item Three hybrid operators :$E_{\downarrow |}$, $@_{||}$ and $E_{| \downarrow }$
  \item A constant symbol $f^{e_{j}}_{l}$  for each $\bm{f}^{e_{j}}_{l}$
  \item A constant symbol $R^{e_{j}}_{k}$ for each $\bm{R}^{e_{j}}_{k}$
  \item Four connectives $\blacktriangle ,\blacktriangleright, \vartriangleright$, and $\vartriangle$.
  \item Auxiliary symbols: $[,],(,)$ and $|$.
\end{itemize}

\paragraph{}
It must be noted that in the alphabet there are neither any kind of variables nor elements to design elements of the set $T$. 

A tuple of symbols $(w^{p,1},...,w^{p,dim(p)})$ is denoted as $v^{p}$.
\newline\newline
The language $PL_{MMPPF}$ has the following three kinds of atomic formulas:
\begin{itemize}
  \item Type I: $[ \circledR| \circledS| o_{i} | P_{0} |,...,|P_{n} ]$ where

   \begin{itemize}
     \item $\circledR \in \{ \varepsilon, h \}$
     \item $\circledS  \in \{ E_{\downarrow |}$, $@_{||}$, $E_{| \downarrow } \}$
     \item $P_{p} \in \mathcal{P} ( H\times W_{p,1}\times\cdots\times W_{p,dim(p)} )$
   \end{itemize}

  \item Type II: $[ \circledR| \circledS | S^{o_{i},p}_{j}| o_{u} ]$  where
    \begin{itemize}
    \item $\circledR \in \{ \varepsilon, h \}$
     \item $\circledS  \in \{ E_{\downarrow |}$, $@_{||}$, $E_{| \downarrow } \}$
  \end{itemize}

  \item Type II: $[ \circledR| \circledS | o_{i}| \vec{a}_{o_{i}}  ]$  where
    \begin{itemize}
    \item $\circledR \in \{ \varepsilon, h \}$
     \item $\circledS  \in \{ E_{\downarrow |}$, $@_{||}$, $E_{| \downarrow } \}$
     \item $\vec{a}_{o_{i}} = (a^{0}_{i},...,a^{n}_{i})$
  \end{itemize}

\end{itemize}
\paragraph{}

Any atomic formula is a well formed formula (wff).\newline\newline

The following rules determine when a wff combined with an atomic formula constitute a wff:
\begin{itemize}

  \item A  wff $\psi=\psi' \blacktriangle p $ and an atomic formula $q$ constitute a wff $\psi \blacktriangle q$ if $\circledS _{p} = \circledS _{q}$ and $\circledS _{p} = \circledS _{q}$

  \item  A  wff $\psi=\psi' \odot \phi $ where $\odot \in \{\blacktriangle ,\blacktriangleright, \vartriangleright\}$ and an atomic formula $q$ constitute a wff $\psi \blacktriangleright q$ if any of the following conditions is fulfilled:
  \begin{itemize}
    \item $\circledS _{p} = E_{\downarrow |}$ and $\circledS_{q} = E_{\downarrow |}$
    \item $\circledS_{p} = E_{\downarrow |}$ and $\circledS_{q} = @_{||}$
    \item $\circledS_{p} = @_{||} $ and $\circledS_{q} = E_{| \downarrow }$
    \item $\circledS_{p} = E_{| \downarrow } $ and $\circledS_{q} = E_{| \downarrow }$
  \end{itemize}

 \item  A  wff $\psi=\psi' \odot p $ where $\odot \in \{\blacktriangle ,\blacktriangleright, \vartriangleright\}$ and an atomic formula $q$ constitute a wff $\psi \vartriangleright q$ if any of the following conditions is fulfilled:
  \begin{itemize}
    \item $\circledS_{p} = E_{\downarrow |}$ and $\circledS_{q} = E_{\downarrow |}$
    \item $\circledS_{p} = E_{\downarrow |}$ and $\circledS_{q} = @_{||}$
    \item $\circledS_{p} = @_{||} $ and $\circledS_{q} = E_{| \downarrow }$
    \item $\circledS_{p} = E_{| \downarrow } $ and $\circledS_{q} = E_{| \downarrow }$
  \end{itemize}

 \item A set of $n$ atomic formulas $p_{1}$,...,$p_{n}$ are a wff $p_{1}\blacktriangle\cdots\blacktriangle p_{n}$ is a wff
 there are not a $p_{x} \equiv[ \circledR| \circledS| o_{i} | P_{0} |,...,|P_{n} ]$ and $p_{y} \equiv[ \circledR'| \circledS'| o_{j} | P'_{0} |,...,|P'_{n} ]$ where $x \neq y$ and $o_{i} = o_{j}$.

   \item A set of $n$ atomic formulas $p_{1}$,...,$p_{n}$ are a wff $p_{1}\blacktriangle\cdots\blacktriangle p_{n}$ is a wff
 there are not a $p_{x} \equiv [ \circledR| \circledS | o_{i}| (a^{0}_{i,k},...,a^{n}_{i,k^{\prime}}) ] $ and $p_{y} \equiv[ \circledR| \circledS | o_{j}| (a^{0}_{j,l},...,a^{n}_{j,l^{\prime}}) ]$ where $o_{i} = o_{j}$.

\end{itemize}

Finally, $\psi \vartriangle \varphi$ is a wff if  $\psi$ and $\varphi$ are wffs.

\subsection{The extended perceptive language of MMPPF }
\label{subsection:epl}
$PL^{*}_{MMPPF}$ is a language of description to MMPPF structures with a higher level of abstraction than $PL_{MMPPF}$. The definition of the $PL^{*}_{MMPPF}$ requires the definition of the alphabets of metainformation $\Sigma_{1}$, $\Sigma_{2}$, $\Sigma_{3}$, $\Sigma_{4}$, $\Pi_{1}$ and $\Pi_{2}$. These alphabets can be assigned to an object from its description in $PL_{MMPPF}$. Thus, the atomic formulas of $PL^{*}_{MMPPF}$ can be built from formulas of $PL_{MMPPF}$. The rules to built a wff of $PL^{*}_{PPFMM}$ are the same that the rules defined in $PL_{PPFMM}$.

\subsubsection*{The metainformation alphabet  $\Sigma_{1}$ }
The metainformation alphabet  $\Sigma_{1}$  has two elements that are denoted by  $0$ and $1$. Thus,
\[
\Sigma_{1}=\{0,1\}
\]

The elements of the metainformation  alphabets are relations. Thus, their definitions are the following:

\begin{itemize}
  \item $\forall o_{i},p,P_{p}\quad 0\langle o_{i},p,P_{p})\rangle \,\longleftrightarrow \,P_{p} = \varnothing$
  \item $\forall o_{i},p,P_{p}\quad 1\langle o_{i},p,P_{p})\rangle \,\longleftrightarrow \,P_{p} \neq \varnothing$
\end{itemize}

Each element of $\Sigma_{1}$ is named  \emph{momentary state}. Using $\Sigma_{1}$, a qualitative state is given  to an object in a moment of time to the $p$-property.  It describes whether an object has a specific quality without naming the specific value. Then, a functor of the language $PL^{*}_{MMPPF}$ is defined, which maps  each object in a moment of time and in relation to a property into the metainformation  alphabet $\Sigma_{1}$. It is denominated \emph{momentary state of the $p$-property }  and denoted by $ms_{p}$, which is defined in the following way:

\[
 ms^{p}(o_{i}, \psi) =
 \begin{cases}
 0, \, \textrm{if} \, P_{p}= \emptyset\\
 1, \, \textrm{if} \, P_{p}\neq \emptyset
 \end{cases}
\]

where $ \psi \equiv [\blacklozenge| \blacksquare | o_{i} | P_{0} |,...,| P_{n} ]$ and $\Box$ a temporal situator.

The following formula can be built using the functor $em_{p}$ :
\[
\psi^{*} \equiv [\blacklozenge| \blacksquare | o_{i} | ms_{0}(o_{i}, \psi) |,...,| ms_{n}(o_{i}, \psi) ]
\]

Thus, $\psi^{*}$, which belongs to $PL^{*}_{MMPPF}$, is built from $\psi$.

\subsection*{The metainformation alphabet   $\Sigma_{2}$ }

The metainformation alphabet  $\Sigma_{2}$   has three elements that are denoted by  $\wr$, $\beta_{1}$ y $\beta_{2}$. Thus,
\[
\Sigma_{2}=\{\wr,\beta_{1},\beta_{2}\}
\]

As the elements of the metainformation alphabets are relations, their definitions are the following:

\begin{itemize}
  \item $\forall P_{p},P'_{p}\quad \beta_{1}\langle P_{p},P'_{p}\rangle \,\longleftrightarrow \,P_{p} = P'_{p}\wedge \phi$
  \item $\forall P_{p},P'_{p}\quad \beta_{2}\langle P_{p},P'_{p}\rangle \,\longleftrightarrow \,P_{p} \neq P'_{p}\wedge \phi$
   \item $\forall P_{p},P'_{p}\quad \wr \langle P_{p},P'_{p}\rangle \,\longleftrightarrow \, P_{p} =\varnothing \vee P'_{p} = \varnothing$
\end{itemize}

where
\[
P_{p}, P'_{p} \in \mathcal{P} ( H\times W_{p,1}\times\cdots\times W_{p,dim(p)} )= \mathcal{P}(H\times V_{p})
\]
and
\[
\phi \equiv  P_{p} \neq \varnothing \wedge P'_{p} \neq \varnothing
\]

It is denominated \emph{temporal state of the $p$-property} to the functor $tsp^{p}$  that is defined in the following way:

\[
tsp^{p}(o_{i},\varphi  )=
\begin{cases}
\beta_{1}, \, \textrm{if} \, P_{p} = P'_{p}\wedge \phi \\
\beta_{2}, \, \textrm{if} \, P_{p} \neq P'_{p}\wedge \phi\\
\wr,       \, \textrm{if} \, ms_{p}(o_{i}, \psi) = 0 \vee ms_{p}(o_{i}, \psi')= 0
\end{cases}
\]
where
\[
\varphi\equiv \psi \blacktriangleright \psi'
\]
and
\[
 \psi \equiv [\blacklozenge| \blacksquare | o_{i} | P_{0} |,...,| P_{dim(p)} ]
\]
\[
\psi' \equiv [\blacklozenge| \blacksquare | o_{i} | P'_{0} |,...,| P'_{dim(p)} ]
\]
and
\[
\phi \equiv   ms^{p}(o_{i}, \psi) \neq 0 \wedge ms^{p}(o_{i}, \varphi)\neq 0
\]

Using the functor $tsp^{p}$, the following formula can be built:
\[
\varphi^{*} \equiv [\blacklozenge| \blacksquare | o_{i} | tsp^{0}(o_{i}, \varphi) |,...,| tsp^{n}(o_{i}, \varphi) ]
\]

Thus, $\varphi^{*}$, which belongs to $PL^{*}_{PPFMM}$, is  built from $\varphi$.

\subsubsection*{The metainformation alphabet  $\Sigma_{3}$ }

The metainformation alphabet  $\Sigma_{3}$  has three elements that are denoted by  $\wr$, $\gamma_{1}$ and $\gamma_{2} $. Thus,
\[
\Sigma_{3}=\{\wr, \gamma_{1}, \gamma_{2} \}
\]

The elements of the metainformation  alphabets are relations. Thus, their definitions are the following:

\begin{itemize}
  \item $\forall P_{p,q},P'_{p,q}\quad \gamma_{1}\langle P_{p,q},P'_{p,q}\rangle \,\longleftrightarrow \,P_{p,q} = P'_{p,q}$
  \item $\forall P_{p,q},P'_{p,q}\quad \gamma_{2}\langle P_{p,q},P'_{p,q}\rangle \,\longleftrightarrow P_{p,q} \neq P'_{p,q}$
   \item $\forall P_{p,q},P'_{p,q}\quad \wr \langle P_{p,q},P'_{p,q}\rangle \,\longleftrightarrow \,   P_{p,q}=\varnothing \vee P'_{p,q} = \varnothing$
\end{itemize}

where
\[P_{p,q} = \{ (h,w_{q})_{i}\} \; h \in H \, w_{k} \in W^{p,q} \textrm{ where } V_{p}= W_{p,1}\times\cdots \times W_{p,q} \times\cdots \times  W_{dim(p)}
\]
\[
P'_{p,q} = \{ (h',w'_{q})_{j}\} \; h' \in H \, w'_{k} \in W^{p,q} \textrm{ where } V_{p}= W_{p,1}\times\cdots \times W_{p,q} \times\cdots \times  W_{dim(p)}
\]

Previously to define the function that assigns elements of $\Sigma_{3}$ to the objects, it is necessary to define a function that assigns to each element essence the value that has in a specific dimension of a specific property. The function is denoted by $c^{p,q}_{j} $ and its definition is the following:
\[
\begin{array}{cccc}
  c^{p,q}_{j} :& H & \rightarrow & W_{p,q} \\
               & h_{i} & \mapsto & w_{q}        \\
\end{array}
\]
where $ g^{p}_{j}(h)_{i} = (w_{1},....,w_{dim(p)}) $

It is denominated \emph{temporal state of the $q$-component of the $p$-property} to the functor $tscp^{p,q}$  that is defined in the following way:

\[
tscp^{p,q}(o_{i},\varphi  )=
\begin{cases}
\gamma_{1}, \, \textrm{if} \,  \phi \wedge \phi' \\
\gamma_{2}, \, \textrm{if} \, \phi  \wedge \phi'' \\
\wr,       \, \textrm{if} \, ms^{p}(o_{i}, \psi) = 0 \vee ms^{p}(o_{i}, \psi')= 0
\end{cases}
\]
where
\[
\varphi\equiv \psi \blacktriangleright \psi'
\]
and
\[
 \psi \equiv [\blacklozenge| \blacksquare | o_{i} | P_{0} |,...,| P_{n} ]
\]
\[
\psi' \equiv [\blacklozenge| \blacksquare| o_{i} |  P'_{0} |,...,| P'_{n} ]
\]
and
\[
\phi \equiv  ms^{p}(o_{i}, \psi) \neq 0 \wedge ms^{p}(o_{i}, \psi')\neq 0
\]
\[
\phi' \equiv \forall h_{i}\in ES_{j}(o_{i})\cap ES_{j'}(o_{i}), c^{*p,q}_{j}(h_{i}) = c^{*p,q}_{j'}(h_{i})
\]
\[
\phi'' \equiv \forall h_{i}\in ES_{j}(o_{i})\cap ES_{j'}(o_{i}), c^{*p,q}_{j}(h_{i}) \neq c^{*p,q}_{j'}(h_{i})
\]
Using the functor $tscp^{p,q}$, the following formula can be built:
\[
\varphi^{*} \equiv [\blacklozenge| \blacksquare | o_{i} | tscp^{p,q}(o_{i}, \varphi) |,...,| tscp^{p,q}(o_{i}, \varphi) ]
\]

Thus, $\varphi^{*}$, which belongs to $PL^{*}_{PPFMM}$, is  built from $\varphi$.

\subsubsection*{The metainformation alphabet  $\Sigma_{4}$ }

The metainformation alphabet  $\Sigma_{4}$  has three elements that are denoted by  $\wr$, $\delta_{1}$ and $\delta_{2} $. Thus,
\[
\Sigma_{4}=\{\wr, \delta_{1}, \delta_{2} \}
\]

The elements of the metainformation  alphabets are relations. Thus, their definitions are the following:

\begin{itemize}
  \item \begin{multline*}
          \forall C_{p,q},C'_{p,q}\quad \delta_{1}\langle  C_{p,q},C'_{p,q}\rangle \,\longleftrightarrow \\
          \forall (h_{j},w_{k})\in C_{p,q}(h_{j'}\, w_{k'}) \in C'_{p,q} h_{j} =  h_{j'}\rightarrow  w_{k} \prec  w_{k'}
        \end{multline*}

  \item  \begin{multline*}
          \forall C_{p,q},C'_{p,q}\quad \delta_{1}\langle  C_{p,q},C'_{p,q}\rangle \,\longleftrightarrow \\
          \forall (h_{j},w_{k})\in C_{p,q}(h_{j'}\, w_{k'}) \in C'_{p,q} h_{j} =  h_{j'}\rightarrow  w_{k} \succ  w_{k'}
        \end{multline*}

   \item  \begin{multline*}\forall C_{p,q},C'_{p,q}\quad \wr \langle C_{p,q},C'_{p,q}\rangle \,\longleftrightarrow \,   C_{q}=\varnothing \vee C'_{q} = \varnothing \\
       \end{multline*}
\end{itemize}

where
\[C_{p,q} = \{ (h_{j},w_{k})\} \; h_{j} \in H \, w_{k} \in W^{p,q} \textrm{ where } V_{p}= W_{1}\times\cdots \times W_{q} \times\cdots \times  W_{dim(p)}
\]
\[
C'_{p,q} = \{ (h_{j'},w_{k'})\} \; h_{j} \in H \, w_{k} \in W^{p,q} \textrm{ where } V_{p}= W_{1}\times\cdots \times W_{q} \times\cdots \times  W_{dim(p)}
\]

It is denominated \emph{ temporal order state of the $q$-component of the $p$-property} to the functor $toscp^{p,q}$  that is defined in the following way:

\[
toscp^{p,q}(o_{i},\varphi  )=
\begin{cases}
\gamma_{1}, \, \textrm{if} \,  \phi \wedge \phi' \\
\gamma_{2}, \, \textrm{if} \, \phi  \wedge \phi'' \\
\wr,       \, \textrm{if} \, ms^{p}(o_{i}, \psi) = 0 \vee ms^{p}(o_{i}, \psi')= 0
\end{cases}
\]
where
\[
\varphi\equiv \psi \blacktriangleright \psi'
\]
and
\[
 \psi \equiv [\blacklozenge| \blacksquare | o_{i} | g^{*0}_{j}(o_{i}) |,...,| g^{*n}_{j}(o_{i}) ]
\]
\[
\psi' \equiv [\blacklozenge| \blacksquare | o_{i} | g^{*0}_{j'}(o_{i}) |,...,| g^{*n}_{j'}(o_{i}) ]
\]
and
\[
\phi \equiv  ms^{p}(o_{i}, \psi) \neq 0 \wedge ms^{p}(o_{i}, \psi')\neq 0
\]
\[
\phi' \equiv \forall h_{i}\in ES_{j}(o_{i})\cap ES_{j'}(o_{i}), c^{*p,q}_{j}(h_{i}) \prec c^{*p,q}_{j'}(h_{i})
\]
\[
\phi'' \equiv \forall h_{i}\in ES_{j}(o_{i})\cap ES_{j'}(o_{i}), c^{*p,q}_{j}(h_{i}) \succ c^{*p,q}_{j'}(h_{i})
\]
Using the functor $tscp^{p,q}$, the following formula can be built:
\[
\varphi^{*} \equiv [\blacklozenge| \blacksquare | o_{i} | toscp^{p,q}(o_{i}, \varphi) |,...,| toscp^{p,q}(o_{i}, \varphi) ]
\]

Thus, $\varphi^{*}$, which belongs to $PL^{*}_{PPFMM}$, is  built from $\varphi$.

\subsubsection*{The metainformation alphabet  $\Pi_{1}$ }

The metainformation alphabet  $\Pi_{1}$  has two elements that are denoted by  $\kappa_{1}$ and $\kappa_{2} $. Thus,
\[
\Pi_{1}=\{ \kappa_{1}, \kappa_{2} \}
\]

It is denominated \emph{ relational state of the $p$-property} to the functor $rs^{p}$  that is defined in the following way:

\[
rs^{p}(o_{i},o_{u},\psi  )=
\begin{cases}
\kappa_{1}, \, \textrm{if} \,  \nexists k \, \psi_{k} \equiv\phi \\
\kappa_{2}, \, \textrm{if} \,  \exists k \, \psi_{k} \equiv\phi \\
\end{cases}
\]
where
\[
\psi \equiv \psi_{1}\blacktriangle \cdots\blacktriangle \psi_{n}
\]
\[
\phi\equiv [\blacklozenge| \blacksquare | S^{o_{i},p}_{j} o_{u} ]
\]

Using the functor $rs^{p}$, the following formula can be built:
\[
\psi^{*} \equiv [\blacklozenge| \blacksquare | o_{i} | o_{u}| rs^{p}(o_{i}, \psi) ]
\]

Thus, $\psi^{*}$, which belongs to $PL^{*}_{PPFMM}$, is  built from $\psi$.

\subsubsection*{The metainformation alphabet  $\Pi_{2}$ }
The metainformation alphabet  $\Pi_{2}$  has four elements that are denoted by  $\tau_{1}$, $\tau_{2}$, $\tau_{3}$ and $\tau_{4}$. Thus,
\[
\Pi_{2}=\{\tau_{1}, \tau_{2},  \tau_{3},  \tau_{4} \}
\]

It is denominated \emph{ temporal relational state of the $p$-property} to the functor $rs^{p}$  that is defined in the following way:

\[
rs^{p}(o_{i},o_{u},\varphi )=
\begin{cases}
\tau_{1}, \, \textrm{if} \,  \nexists k, k'\; \psi_{k} \equiv\phi \, \wedge  \, \psi'_{k} \equiv \phi'\\
\tau_{2}, \, \textrm{if} \,  \nexists k \exists k' \, \psi_{k} \equiv\phi   \wedge  \, \psi'_{k} \equiv \phi'\\
\tau_{3}, \, \textrm{if} \,  \exists k \nexists k' \, \psi_{k} \equiv\phi  \wedge  \, \psi'_{k} \equiv \phi'\\
\tau_{4}, \, \textrm{if} \,  \exists k, k' \, \psi_{k} \equiv\phi    \wedge  \, \psi'_{k} \equiv \phi'\\
\end{cases}
\]
where
\[
\varphi \equiv \psi_{1}\blacktriangle \cdots\blacktriangle \psi_{n}\blacktriangleright \psi'\cdots\blacktriangle \psi_{n}
\]
\[
\phi\equiv [\blacklozenge| \blacksquare| S^{o_{i},p}_{j} o_{u} ]
\]
\[
\phi'\equiv [\blacklozenge| \blacksquare | S^{o_{i},p}_{j'} o_{u} ]
\]
Using the functor $rs^{p}$, the following formula can be built:
\[
\psi^{*} \equiv [\blacklozenge| \blacksquare | o_{i} | o_{u}| rs^{p}(o_{i}, \varphi) ]
\]

Thus, $\psi^{*}$, which belongs to $PL^{*}_{PPFMM}$, is  built from $\psi$.

\subsection{The categorical language of MMPPF }

$CL_{MMPPF}$ is at the top of the hierarchy of languages. It uses the metainformation alphabets of $PL^{*}_{MMPPF}$, but the atomic formulas qualify the intervals of time. It is abstracted from the duration of the time interval. There are four kinds of atomic formulas as are in $PL_{MMPPF}$ . The atomic formulas can be created with elements that belong to sets called categories. There are five categories: \emph{objects}, \emph{patterns of objects}, \emph{conditions of reality}, \emph{temporal situators}, and \emph{registers of states}.

\begin{itemize}
  \item Objects $\mathcal{C}_{O}= \{ o_{1},...,o_{z} \}$
  \item Patterns of objects $\mathcal{C}_{PO}= \{ \lambda o.\varphi_{i},... \}$ where $\varphi_{i}$ is a formula that is true if the object fulfills a specific feature.
  \item Conditions of Reality $\mathcal{C}_{CR}= \{ \varepsilon,h \}$
  \item Temporal situators $\mathcal{C}_{TS}= \{ \downarrow|,||,|\downarrow \}$
  \item Registers of states $\mathcal{C}_{RS}$ whose elements are functions that produce atomic formulas. There are three kinds of functions.

    \begin{itemize}
                   \item $\lambda tp \lambda o_{i} [tp| o_{i}|p| x ] \qquad x \in \sum_{1}\cup \sum_{2}$
                   \item $\lambda tp \lambda o_{i} [tp| o_{i}|p|q| x ] \qquad x \in \sum_{3}\cup \sum_{4}$
                   \item $\lambda tp \lambda o_{i} [tp| o_{i}|o_{u}| x ] \qquad x \in \prod_{2}$
  \end{itemize}

  where $p$ determines a property, $q$ determines a dimension of a property, and $tp$ a  determines temporal positioner.
\end{itemize}

Each element of a category is named \emph{atom}. An atomic well-formed formula (wff) of  $CL_{MMPPF}$ can be generated by applying  $\beta$-reduction  to the atoms according to their types. The compound formulas of  $CL_{MMPPF}$  have $\blacktriangle ,\blacktriangleright, \vartriangleright$, and $\vartriangle$ as connective. However, the meaning of the connectives is different in  $CL_{MMPPF}$ from $PL_{MMPPF}$ and $PL^{*}_{MMPPF}$ because atomic formulas are not about moments of time but intervals of time. Thus, the connectives relate intervals of time.

\section{Semantics}
We have defined a class of mathematical structures and three formal languages to describe structures of that class. We need to know when a formula describe rightly a structure. Since, the formal languages is a hierarchy, the true of a formula of a language can be derived from the true of a formula of a language of a lower level.   Thus, to provide a semantics for the hierarchy of languages, firstly, it must be defined a satisfiability for each formula of $PL_{MMPPF}$.  To do that we need an interpretation to each symbol that composes each atomic formula. Because the alphabet of $PL_{MMPPF}$ does not have variables we do not need an assignation function. Given a $\bm{r}_{e }= (\circledR,\circledS,t,\bm{e}_{i}) $ , the interpretation function $\mathcal{I}^{\bm{r}_{e}}$ fulfills  the following $ \mathcal{I}^{\bm{e}_{i}}  \subset\mathcal{I}^{\bm{r}_{e }}$. Thus, for example, the following is fulfilled:

\[
\mathcal{I}^{\bm{r}_{j}}(S^{o_{i}}) = \mathcal{I}^{\bm{e}_{i}}(S^{o_{i},p})  = \bm{S}^{\bm{o}_{i},p}
\]

We will use the letters: $p$ to designate atomic formulas of type I and II, $i$ to designate atomic formulas of type III, and $\varphi,\psi,...$ to designate formulas of $PL_{MMPPF}$.

\subsection{The satisfiability relation of $PL_{MMPPF}$}
The satisfiability relation of a formula of $PL_{MMPPF}$ with $\blacktriangleright$ or $\rhd$  is not only about the satisfiability of the two sides of the connective in theirs moments of time also those connectives mean that there is a transformation from one to the other. For example, if $p$ is true in $\bm{m}_{\bm{t}}$ and $q$ is true in $\bm{m}_{\bm{t}+1}$ we cannot ensure that $p\blacktriangleright q$ is true. Only if there is a path from any of the realities in which $p$ is true to any of the realities of the next moment of time in which $q$ is true, then $p\blacktriangleright q$ is true. Due to that fact, the definition of satisfiability relation  to the connectives $\blacktriangleright$ and $\rhd$ is complex.

Being $\mathfrak{M}$ a MMPPF structure, the satisfiability relation $\Vdash$  to the formulas of $PL_{MMPPF}$ is defined as follows:

\begin{multline*}
\mathfrak{M},\mathcal{P}_{\bm{t}} \Vdash \phi  \textrm{ iff } \mathfrak{M},\mathcal{P}_{\bm{t}},\min(\bm{T}) \Vdash^{^{\leq}} \phi\\
\end{multline*}

\begin{multline*}
\mathfrak{M},\mathcal{P}_{\bm{t}}, \bm{t}'  \Vdash^{^{\leq}} p_{1}\blacktriangle \cdots\blacktriangle  p_{n} \blacktriangle i_{1} \blacktriangle \cdots\blacktriangle  i_{n'} \blacktriangleright \varphi   \textrm{ iff } \\
\textrm{ iff there exist } \bm{t}'', \bm{t}' \leq t''\textrm{ and } RI^{t''} \neq \varnothing  \textrm{ and }\mathfrak{M},\mathcal{P}_{\bm{t}},\bm{t}''', RI  \Vdash^{=} \varphi   \textrm{ where }
\bm{t}'''=\bm{t}''+1 \textrm{ and }\\ \textrm{ and } RI =\{ \langle \bm{r}_{x},\bm{I}' \rangle: \bm{r}_{x} \in \bm{m}_{t'} \textrm{ and } \mathfrak{M},\mathcal{P}_{\bm{t}},\bm{r}_{x}\Vdash p_{1}\blacktriangle \cdots\blacktriangle  p_{n} \textrm{ and } \mathfrak{M},\mathcal{P}_{\bm{t}}, \bm{r}_{x} \Vdash i_{1} \blacktriangle \cdots\blacktriangle i_{n'} \textrm{ and } \\ \textrm{ and }  \bm{I}'=\{\vec{\bm{i}}: \vec{\bm{i}} \in \bm{I}_{x} \textrm{ and } \pi^{\bm{o}_{i_{1}}}(\vec{\bm{i}})= \mathcal{I}^{\bm{r}_{x}}(\vec{a}_{o_{i_{1}}}) \textrm{ and }... \textrm{ and }  \pi^{\bm{o}_{i_{n'}}}(\vec{i}) = \mathcal{I}^{r_{x}}(\vec{a}_{o_{i_{n'}}}) \} \\
\end{multline*}

\begin{multline*}
\mathfrak{M},\mathcal{P}_{\bm{t}}, \bm{t}', RI \Vdash^{^{=}} p_{1}\blacktriangle \cdots\blacktriangle  p_{n} \blacktriangle i_{1} \blacktriangle \cdots\blacktriangle  i_{n'} \blacktriangleright \varphi \textrm{ iff } \\
\textrm{ iff } R'' \neq \varnothing    \textrm{ and } \mathfrak{M},\mathcal{P}_{\bm{t}},t'', RI'  \Vdash \varphi \textrm{ where }\\
\textrm{ where }  R'' =\{  \bm{r}_{v} : \bm{r}_{v}\in R', (\bm{r}_{u},\bm{I}') \in RI, \bm{i}_{w}\in \bm{I}' \textrm{ where } \bm{\&}(\bm{e}_{x},\bm{i}_{w})=\bm{e}_{y} \} \neq \varnothing \textrm{ and }\\
  \textrm{ and } \bm{t}''=t'+1 \textrm{ and } \\
  \textrm{ and } R' =\{ \bm{r}_{i} : \bm{r}_{i}\in \bm{m}_{t'} \textrm{ and } \mathfrak{M},\mathcal{P}_{\bm{t}}, \bm{r}_{i}\Vdash p_{1}\blacktriangle \cdots\blacktriangle  p_{n} \textrm{ and } \mathfrak{M},\mathcal{P}_{\bm{t}}, \bm{r}_{i} \Vdash i_{1} \blacktriangle \cdots\blacktriangle i_{n'}  \}  \textrm{ and } \\ \textrm{ and }
  \bm{r}_{u}= (t_{a},\blacklozenge,\blacksquare,\bm{e}_{x}) \textrm{ and }  \bm{r}_{v}=(t',\blacklozenge',\blacksquare',\bm{e}_{y}) \textrm{ and } \\
\textrm{ and } \textrm{ and }   RI' =\{ \langle \bm{r}_{v},\bm{I}''\rangle : \bm{r}_{v}\in R'' \textrm{ and } \bm{I}''=\{\vec{\bm{i}}: \vec{\bm{i}} \in \bm{I}_{e_{x}} \pi^{\bm{o}_{i_{1}}}(\vec{\bm{i}})= \mathcal{I}^{\bm{r}_{x}}(\vec{a_{o_{i_{1}}}}) \textrm{ and } ...\\... \textrm{ and }  \pi^{\bm{o}_{i_{n'}}}(\vec{\bm{i}})= \mathcal{I}^{\bm{r}_{x}}(\vec{a}_{o_{i_{n'}}}) \}\} \\
\end{multline*}

\begin{multline*}
\mathfrak{M},\mathcal{P}_{\bm{t}},\bm{r}_{e} \Vdash p_{1}\blacktriangle \cdots\blacktriangle  p_{n}
\textrm{ iff } \mathfrak{M},\mathcal{P}_{\bm{t}},  \bm{r}_{e} \Vdash p_{1}  \textrm{ and } ...\\
 ... \textrm{ and } \mathfrak{M},\mathcal{P}_{\bm{t}}, \bm{r}_{e}  \Vdash p_{n}  \\
\end{multline*}

\begin{multline*}
\mathfrak{M},\mathcal{P}_{\bm{t}},\bm{r}_{e} \Vdash i_{1}\blacktriangle \cdots\blacktriangle  i_{n'}
\textrm{ iff } \mathfrak{M},\mathcal{P}_{\bm{t}},  \bm{r}_{e} \Vdash i_{1}  \textrm{ and } ...\\
 ... \textrm{ and } \mathfrak{M},\mathcal{P}_{\bm{t}}, \bm{r}_{e}  \Vdash i_{n'}  \\
\end{multline*}


\begin{multline*}
\mathfrak{M},\mathcal{P}_{\bm{t}}, \bm{t}'  \Vdash^{^{\leq}} p_{1}\blacktriangle \cdots\blacktriangle  p_{n}\blacktriangle i_{1} \blacktriangle \cdots\blacktriangle  i_{n'} \rhd \varphi   \textrm{ iff there exists } \\
\textrm{ iff there exists } t'' \textrm{ and } \bm{t}' \leq \bm{t}''  \textrm{ and } RI^{\bm{t}''}\neq\varnothing
\textrm{ and }\mathfrak{M},\mathcal{P}_{\bm{t}},\bm{t}''' RI^{t''}  \Vdash^{^{<}} \varphi   \textrm{ where }\\
\textrm{ where } \bm{t}'''=\bm{t}''+1 \textrm{ and }\\ \textrm{ and }  RI^{t''} =\{ \langle\bm{r}_{e},\bm{I}' \rangle: \bm{r}_{e} \in \bm{m}_{t''} \textrm{ and } \mathfrak{M},\mathcal{P}_{\bm{t}},\bm{r}_{e}\Vdash p_{1}\blacktriangle \cdots\blacktriangle  p_{n} \textrm{ and } \mathfrak{M},\mathcal{P}_{\bm{t}}, \bm{r}_{e} \Vdash i_{1} \blacktriangle \cdots\blacktriangle i_{n'} \textrm{ and } \\  \textrm{ and } \bm{I}'=\{\vec{\bm{i}}: \vec{\bm{i}} \in \bm{I}_{x} \pi^{o_{i_{1}}}(\vec{\bm{i}})= \mathcal{I}^{\bm{r}_{x}}(\vec{a_{o_{i_{1}}}})\textrm{ and }... \textrm{ and }  \pi^{\bm{o}_{i_{n'}}}(\vec{\bm{i}})= \mathcal{I}^{\bm{r}_{v}}(\vec{a}_{o_{i_{n'}}}) \}  \}
\end{multline*}

\begin{multline*}
\mathfrak{M},\mathcal{P}_{\bm{t}}, \bm{t}', RI \Vdash^{^{=}} p_{1}\blacktriangle \cdots\blacktriangle  p_{n} \blacktriangle i_{1} \blacktriangle \cdots\blacktriangle  i_{n'} \rhd \varphi \textrm{ iff } \\
\textrm{ iff } RI' \neq \varnothing    \textrm{ and }
\textrm{ and } \mathfrak{M},\mathcal{P}_{\bm{t}},t'', RI'  \Vdash^{^{\leq}} \varphi  \textrm{ where } \bm{t}''=\bm{t}'+1  \textrm{ and }\\
R' =\{ \bm{r}_{e} : \bm{r}_{e}\in \bm{m}_{t'} \textrm{ and } \mathfrak{M},\mathcal{P}_{\bm{t}}, \bm{r}_{e}\Vdash p_{1}\blacktriangle \cdots\blacktriangle  p_{n}  \textrm{ and } \mathfrak{M},\mathcal{P}_{\bm{t}}, r_{e} \Vdash i_{1} \blacktriangle \cdots\blacktriangle i_{n'}  \}\\
\textrm{ and }  RI' =\{ (\bm{r}_{v},\bm{I}'') :\bm{r}_{v}\in R',(\bm{r}_{u},\bm{I}')\in RI, \bm{i}_{w}\in \bm{I}_{e_{x}} \textrm{ where } \bm{\&}(\bm{e}_{x},\bm{i}_{w})=\bm{e}_{y} \textrm{ and } \\
\textrm{ and } \bm{I}''=\{\vec{\bm{i}}: \vec{\bm{i}} \in \bm{I}_{\bm{e}_{y}} \pi^{\bm{o}_{i_{1}}}(\vec{\bm{i}})= \mathcal{I}^{\bm{r}_{v}}(\vec{a_{o_{i_{1}}}}) \textrm{ and }... \textrm{ and }  \pi^{\bm{o}_{i_{n'}}}(\vec{\bm{i}})= \mathcal{I}^{\bm{r}_{v}}(\vec{a}_{o_{i_{n'}}}) \} \}  \textrm{ and } \\ \textrm{ and } \bm{r}_{v}=(\bm{t}',\blacklozenge',\blacksquare',\bm{e}_{y}) \textrm{ and } (\bm{r}_{u}=  (\bm{t}_{a},\blacklozenge,\blacksquare,\bm{e}_{x})\\
\end{multline*}

\begin{multline*}
\mathfrak{M},\mathcal{P}_{\bm{t}}, \bm{t}', RI \Vdash^{<} p_{1}\blacktriangle \cdots\blacktriangle  p_{n} \blacktriangle i_{1} \blacktriangle \cdots\blacktriangle  i_{n'} \rhd \varphi \textrm{ iff there exists }\\
\textrm{ iff there exists } \bm{t}'', \textrm{ and } \bm{t}' < \bm{t}'' \textrm{ and }  R'\neq \varnothing   \textrm{ and } \mathfrak{M},\mathcal{P}_{\bm{t}},\bm{t}''', RI_{\bm{t}''}'  \Vdash^{^{<}} \varphi \textrm{ where} \\
\textrm{ where }  RI_{\bm{t}''}' =\{ \bm{r}_{v} : \bm{r}_{v}\in \bm{m}_{\bm{t}''} \textrm{ and }  \mathfrak{M},\mathcal{P}_{\bm{t}}, \bm{r}_{v}\Vdash p_{1}\blacktriangle \cdots\blacktriangle  p_{n}   \textrm{ and } \mathfrak{M},\mathcal{P}_{\bm{t}}, \bm{r}_{v} \Vdash i_{1} \blacktriangle \cdots\blacktriangle i_{n'} \textrm{ and there} \\
\textrm{ and there exist an  } (\bm{i}_{1},...,\bm{i}_{m})  \textrm{ where }\bm{t}''-\bm{t}'=m \textrm{ and } \bm{i}_{1}\in \bm{I}' \textrm{ and } (\bm{r}_{u},\bm{I}')\in RI \textrm{ and }\\ \textrm{ and } \bm{\&}(\&(...\bm{\&}(\bm{e}_{x},\bm{i}_{1}),...),\bm{i}_{m-1}),\bm{i}_{m})=\bm{e}_{y} \} \\
\textrm{ and }  \bm{t}'''=\bm{t}''+1  \textrm{ and }  \bm{r}_{v}=(\bm{t}',\blacklozenge',\blacksquare',\bm{e}_{y}) \textrm{ and }\bm{r}_{u} = (\bm{t}_{a},\blacklozenge,\blacksquare,\bm{e}_{x})
\end{multline*}

\begin{multline*}
\mathfrak{M},\mathcal{P}_{\bm{t}},\bm{t}'  \Vdash \varphi\vartriangle \psi  \textrm{ iff  } \mathfrak{M},\mathcal{P}_{\bm{t}}, \bm{t}' \Vdash \varphi  \textrm{ and } \mathfrak{M},\mathcal{P}_{\bm{t}},\min(\bm{T})\Vdash \psi \\
\end{multline*}

\begin{multline*}
\mathfrak{M},\mathcal{P}_{\bm{t}},\bm{r}_{e}= (\bm{t}',\blacklozenge ,\blacksquare, \bm{e}_{x} ) \Vdash [ \circledR| \circledS| o_{i} | P_{0} |,...,|P_{n} ]\textrm{ iff } \\ \textrm{ iff } \bm{r}_{e} \in \bm{m}_{t'} \in M_{\mathcal{P}_{\bm{t}}}   \textrm{ and } \\
\textrm{ and } \mathcal{I}^{\bm{r}_{e}}(\circledR) =\blacklozenge \textrm{ and } \mathcal{I}^{\bm{r}_{e}}(\circledS) = \blacksquare
\textrm{ and for all }P_{p} \textrm{, }\mathcal{I}^{\bm{r}_{e}}(P_{p}) = g^{*p}_{j}(\mathcal{I}^{\bm{r}_{e}}(o_{i}))
\end{multline*}

\begin{multline*}
\mathfrak{M},\mathcal{P}_{\bm{t}},\bm{r}_{e}= (t',\blacklozenge ,\blacksquare, \bm{e}_{x} ) \Vdash [ \circledR| \circledS| S^{o^{i},p}| o_{u} ]\textrm{ iff } \\ \textrm{ iff  } \bm{r}_{e} \in \bm{m}_{t'} \in M_{\mathcal{P}_{\bm{t}}} \textrm{ and  } \\
\textrm{ and  } \mathcal{I}^{\bm{r}_{e}}(\circledR) =\blacklozenge \textrm{ and } \mathcal{I}^{r_{e}}(\circledS) = \blacksquare \textrm{ and } \mathcal{I}^{\bm{r}_{e}}( o_{u}) \in \mathcal{I}^{r_{e}}(S^{o^{i},p})
\end{multline*}

\begin{multline*}
 \mathfrak{M},\mathcal{P}_{\bm{t}},\bm{r}_{e}= (t',\blacklozenge ,\blacksquare, e_{x} ) \Vdash [ \varepsilon| \circledS| o_{i}| (a^{0}_{i,k},...,a^{n}_{i,k^{\prime}}) ] \textrm{ where }
   \bm{t}' \leq \bm{t}  \textrm{ iff } \\ \textrm{ iff } \bm{r}_{e} \in \bm{m}_{t'}  \in M_{\mathcal{P}_{\bm{t}}} \textrm{ and } \quad \funcii (\bm{t}') = (\mathcal{I}^{\bm{r}_{e}} (a^{0}_{i}),..., \mathcal{I}^{\bm{r}_{e}} (a^{n}_{i}) ) \textrm{ and }
 \\ \textrm{ and } \mathcal{I}^{\bm{r}_{e}} (\circledS) = \blacksquare' \textrm{ where } \bm{r}_{e} = (t',\blacklozenge' ,\blacksquare', \bm{e}_{x'} )
\end{multline*}

\begin{multline*}
 \mathfrak{M},\mathcal{P}_{\bm{t}},\bm{r}_{e}= (t',\blacklozenge ,\blacksquare, e_{x} ) \Vdash [ \varepsilon| \circledS| o_{i}| (a^{0}_{i,k},...,a^{n}_{i,k^{\prime}}) ] \textrm{ where }
   \bm{t}' > \bm{t}  \textrm{ iff } \\ \textrm{ iff } \bm{r}_{e} \in \bm{m}_{t'}  \in M_{\mathcal{P}_{\bm{t}}} \textrm{ and } \mathcal{I}^{\bm{r}_{e}} (a^{0}_{i})\in \theta^{0}_{x}( \mathcal{I}^{\bm{r}_{e}}(o_{i}) ) \textrm{ and },..., \textrm{ and } \mathcal{I}^{\bm{r}_{e}} (a^{n}_{i})\in \theta^{n}_{x}( \mathcal{I}^{\bm{r}_{e}}(o_{i}) ) \textrm{ and }
 \\ \textrm{ and } \mathcal{I}^{\bm{r}_{e}} (\circledS) = \blacksquare' \textrm{ where } \bm{r}_{x} = (t',\blacklozenge' ,\blacksquare', \bm{e}_{x'} )
\end{multline*}

\begin{multline*}
 \mathfrak{M},\mathcal{P}_{\bm{t}},\bm{r}_{e}= (t',\blacklozenge ,\blacksquare, e_{x} ) \Vdash [ h| \circledS| o_{i}| (a^{0}_{i},...,a^{n}_{i}) ]\textrm{ iff } \\ \textrm{ iff  }  \bm{r}_{e} \in \bm{m}_{t'}  \in M_{\mathcal{P}_{\bm{t}}} \textrm{ and }
  \mathcal{I}^{\bm{r}_{e}} (a^{0}_{i})\in \theta^{0}_{x}( \mathcal{I}^{\bm{r}_{e}}(o_{i}) ) \textrm{ and },..., \textrm{ and } \mathcal{I}^{\bm{r}_{e}} (a^{n}_{i})\in \theta^{n}_{x}( \mathcal{I}^{\bm{r}_{e}}(o_{i}) ) \textrm{ and } \\
\textrm{ and } \mathcal{I}^{\bm{r}_{e}} (\circledS) = \blacksquare' \textrm{ where } \bm{r}_{e} = (t',\blacklozenge' ,\blacksquare', \bm{e}_{x'} )
\end{multline*}

The connectives $\blacktriangle$ and $\vartriangle$ are commutative  while $\blacktriangleright$ and $\triangleright$ are noncommutative.

\subsection{The satisfiability relation of $PL^{*}_{MMPPF}$ }

Once the satisfiability relation of $PL_{MMPPF}$ has been defined, using a translation function we define the  satisfiability relation of $PL^{*}_{MMPPF}$. The translation function is $Tr_{1}$ defined in the following way
\[
Tr_{1}: PL_{MMPPF} \longrightarrow PL^{*}_{MMPPF}
\]
The translation function $Tr_{1}$ uses the functions defined in the subsection \ref{subsection:epl}.
Thus, we define the  satisfiability relation of $PL^{*}_{MMPPF}$ as follows:

\begin{multline*}
\mathfrak{M},\mathcal{P}_{\bm{t}} \Vdash \phi  \textrm{ iff there exists an }\varphi \in L_{MMPPF} \textrm{ where }
Tr_{1}(\varphi)=\phi \textrm{ and }
\mathfrak{ M },\mathcal{ P }_{t},\min( T ) \Vdash^{^{\leq}} \varphi\\
\end{multline*}

\subsection{The satisfiability relation of $LC_{MMPPF}$ and the recognizer grammars of true conditions }

Finally, once defined the  satisfiability relation of $PL^{*}_{MMPPF}$ we can define the satisfiability relation of $LC_{MMPPF}$ with the same method. Using a translation function, $Tr_{2}$,  defined in the following way:
\[
Tr_{2}: PL^{*}_{MMPPF} \longrightarrow LC_{MMPPF}
\]
Thus, we define the  satisfiability relation of $LC_{MMPPF}$ in the following way:
\begin{multline*}
\mathfrak{M},\mathcal{P}_{\bm{t}} \Vdash \phi  \textrm{ iff there exists an }\varphi \in PL^{*}_{MMPPF} \textrm{ where }
Tr (\varphi)=\phi \textrm{ and }
\mathfrak{ M },\mathcal{ P }_{t},\min( T ) \Vdash^{^{\leq}} \varphi\\
\end{multline*}

The important issue about the satisfiability relation of $LC_{MMPPF}$  is how define $Tr_{2}$ because building a formula of $CL_{MMPPF}$ from $PL^{*}_{MMPPF}$ is a much more complicated task because we have many formulas of $PL^{*}_{MMPPF}$  that must be associated with only one formula of $CL_{MMPPF}$. This is caused because $CL_{MMPPF}$ bringing the qualification on time, so there is not a specific duration associated with a formula of $CL_{MMPPF}$. Thus, we need a mechanism that allows  many-1 translations. It involves recognizing formulas of $PL^{*}_{MMPPF}$ and generating formulas of $CL_{MMPPF}$.

A formal grammar is a mechanism that allows recognition of an infinite number of expressions with a finite number of rules. However, this is not sufficient because we need the capacity to associate expressions of one language with expressions of another language. Fortunately, there is a mechanism that has been studied that has both features, syntax directed translation scheme(SDTS) \cite{Aho1969}. The interesting issue of the SDTSs is that they permit many-1 translations, which is the feature needed to resolve the problem. We use the SDTS to define a mechanism named \emph{recognizer grammar of true conditions}(RGTC). The RGTC is the mechanism that the CTTC uses to build a formula of $CL_{MMPPF}$ from $PL^{*}_{MMPPF}$. Thus, we begin with defining translation and the SDTS.

A translation is a subset of $\Sigma^{*} \times \Upsilon^{*}$ for finite alphabets $\Sigma$ and $\Upsilon$. An SDTS is a system that generalizes the notion of a context free grammar to generate a translation. It is denoted $G = (N, \Sigma, \Upsilon, R, S)$, where $N$, $\Sigma$, $\Upsilon$ are finite sets of \emph{nonterminal symbols}, \emph{input symbols}, and \emph{output symbols}, respectively. $N$ is disjoint from $\Sigma \cup \Upsilon$. $S \in N$, is the start symbol. $R$ is the finite set of rules.
A rule is an object $A \longrightarrow (\alpha, \gamma,\Pi)$, where $A$ is a nonterminal symbol, where $\alpha \in (N \cup \Sigma)^{*}$, $\gamma \in (N \cup \Upsilon)^{*}$ and $\Pi$ is a permutation.

An SDTS $G = (V, \Sigma, \Upsilon, R, S)$ is simple if for all $A \longrightarrow (\alpha, \gamma,\Pi) \in R$, $\Pi$ is an identity permutation (i.e., $\Pi(i)= i$ for all $i$.). Since the permutation portion of a rule is irrelevant for a simple SDTS, it is deleted from all rules.

An RGTC is an SDTS with production rules that also have a set of assignments.  The RGTC is a simple SDTS whose right sides of its production rules have at most one nonterminal symbol in $\alpha$ and $\gamma$. This is because the translations from $PL^{*}_{MMPPF}$ to $CL_{MMPPF}$  can be done sequentially because each formula of  $PL^{*}_{MMPPF}$  is a sequence of atomic formulas that describes an object in a temporal way. This choice is because psychological time flows in only one direction. However, the RGTC is not only a simple SDTS becaue each rule also contains a set of assignments.  This happens because the elements generated in the translation are $\lambda$-terms. Thus, $\gamma$ allows a set of assignments $\Lambda $. The general form for the rule of an RGTC is as follows:
\[
A \longrightarrow (\alpha, \gamma,\Lambda )
\]

So, given $\gamma= \lambda x.M $ and $\Lambda ={ x:=c} $, the following process is done:
\[
A \, \rightarrow \,  \lambda x.M \{ x:=c\} \, \rightarrow_{\beta} M(c)
\]

The first transformation is generated by the rule of the RGTC, and the second is done by $\beta$-reduction.

The important issue of the assignations is that they can delay a decision about the translation until move forward in the input formula.

\bibliographystyle{splncs03}
\bibliography{BibCTTC}

\end{document}